\documentclass{sig-alternate}
\usepackage[utf8]{inputenc}
\usepackage[T1]{fontenc}
\usepackage{lmodern}
\usepackage{tabularx}
\usepackage{framed}
\usepackage{subfigure}
\usepackage{color}
\usepackage[colorlinks]{myhyperref}

\usepackage{listings}
\usepackage{textcomp}
\usepackage{textpos}
\usepackage{comment}

\newenvironment{definition}[1][Definition]{\begin{trivlist}
\item[\hskip \labelsep {\bfseries #1}]}{\end{trivlist}}

\usepackage[figure,boxed,ruled,vlined]{algorithm2e}
\usepackage{algpseudocode}

\algrenewcommand\algorithmicdo{\textbf{v\’egezd el}}

\newlength\mylen

\begin{document}

\title{Reasoning and Improving on Software Resilience against Unanticipated Exceptions}

\numberofauthors{3}

\author{
Benoit Cornu,
Lionel Seinturier,
and
Martin Monperrus\\
University of Lille \& INRIA\\
\\
Technical report, 2013
}

\maketitle
\vspace{-1.5cm}
\begin{abstract}
In software, there are the errors anticipated at specification and design time, those encountered at development and testing time, and those that happen in production mode yet never anticipated.
In this paper, we aim at reasoning on the ability of software to correctly handle unanticipated exceptions. 
We propose an algorithm, called short-circuit testing, which injects exceptions during test suite execution so as to simulate unanticipated errors.
This algorithm collects data that is used as input for verifying two formal exception contracts that capture two resilience properties.
Our evaluation on 9 test suites, with 78\% line coverage in average, analyzes 241 executed catch blocks, shows that 101 of them expose resilience properties and that 84 can be transformed to be more resilient.
\end{abstract}

\section{Introduction}
At Fukushima's power plant, the anticipated maximum tsunami height was 
5.6m \cite{fukushima}. On March 11, 2011, the highest waves struck at 15m. 
In software, there are the errors anticipated at specification and design time, those encountered at development and testing time, and those that happen in production mode yet never anticipated, as  Fukushima's tsunami.
In this paper, we aim at reasoning on the ability of software to correctly handle unanticipated errors. 

We call this ability ``software resilience''.
It is complementary to the concepts of robustness and fault tolerance~\cite{laprie2008}.
Software robustness emphasizes that the system under study resists to incorrect input data (whether malicious or buggy).
Fault tolerance can have a wide acceptation \cite{avizienis2004basic}, but is mostly  associated with hardware faults.
Software resilience conveys the notion of risks from unanticipated errors (whether environmental or internal) at the software level.

We focus on the resilience against exceptions.
We define it as its ability to reenter a correct state when an exception occurs.
Exceptions are programming language constructs for handling errors \cite{JLS3:05:Sun}.
Exceptions are widely used in practice \cite{cabral2007exception}.
To us, the resilience against exceptions is the ability to correctly handle exceptions that were never foreseen at specification time neither encountered during development.
Our motivation is to help the developers to understand and improve the resilience of their applications.

This sets a three-point research agenda: (RQ\#1) What does it mean to specify anticipated exceptions? (RQ\#2) How to characterize and measure resilience against unanticipated exceptions? (RQ\#3) How to put this knowledge in action to improve the resilience?

In this paper, 
we consider the test suites as specification, since they are available in many existing programs and are pragmatic approximations of idealized specifications\cite{staats2011programs}. 
Test suites specifiy exceptions:
we show in this paper that between 4\% and 26\% of test cases (in a dataset of 9 well-tested open-source applications) specify runtime states that trigger exceptions.

We then define two contracts on the programming language construct ``try-catch'' that capture two facets of software resilience.
We describe two formal criteria and an algorithm, called ``short-circuit testing''. Together they assess whether a try-catch block satisfies those contracts.
Short-circuit testing consists in injecting exceptions in try blocks.
By showing that the test suite still passes under exception injection, 
short-circuit testing uncovers resilience as defined by the ability of recovering to unanticipated problems.
We show that our approach finds much resilient code as formalized by our exception contracts: in our dataset of 9 open-source applications, 92 try-catch blocks expose such resilience.

Finally, we use the knowledge on resilience obtained with short-circuit testing to replace in a catch the caught type by one of its super-type.
This source code transformation is considered as correct if the test suite continues to pass.
By enabling catch blocks to correctly handle more types of exception (w.r.t. the specification), the code is more capable of handling unanticipated exceptions.

\begin{framed}
Our approach helps the developers to be aware of what part of their code is resilient, and to automatically recommend modifications of catch blocks to improve the application's resilience. 
\end{framed}

\newpage
To sum up, our contributions are:
\begin{itemize}
\item A characterization and empirical study of specification of exception handling in test suites,
\item A definition and formalization of two contracts on try-catch blocks,
\item An algorithm and four predicates to verify whether a try-catch satisfies those contracts,
\item A source code transformation to improve resilience against exceptions,
\item An empirical evaluation on 9 open-source software applications showing that there exists resilient try-catch blocks in practice.
\end{itemize}

\section{Characterizing the Specifica-\\tion of Error-Handling in Test Suites}
\label{sec:specification-error-handling}

A test suite is a collection of test cases where each test case contains a set of assertions \cite{beizer2003software}.
The assertions specify what the software is meant to do.
Hence, in the rest of this paper, we consider that a test suite is a specification\footnote{Conversely, when we use the term ``specification'', we refer to the test suite as the specification (even if they are an approximation of an idealized specification \cite{staats2011programs})}.

For instance, ``assert(3, division(15,5))'' specifies that the result of the division of 15 by 5 should be~3.
But when software is in the wild, it may be used with incorrect input or encounter internal errors \cite{zhang2012amplifying}.
For instance, what if one calls ``division(15,0)''?
Consequently, a test suite may also encode what a software package does outside standard usage.
For instance, one may specify that ``division(15,0)'' should throw an exception "Division by zero not possible".

The exceptions that are thrown during test suite execution are anticipated.
The assertions evaluated after caught exceptions specify that the exception-handling code has worked as expected.
Our key insight is that we can use those assertions as an oracle for the resilience capabilities against unanticipated errors (Section \ref{sec:verifying}).
That's why we first need to know how and to what extent error-handling is specified before going further.

We now present a characterization and empirical study of how exception-handling is specified in test suites.
To our knowledge, there is no such empirical study in the literature.

\subsection{Definition of Three New Types of Test Case}
The classical way of analyzing the execution of test suites is to separate passing \textcolor{green}{``green test cases''} and failing \textcolor{red}{ ``red test cases''} \footnote{those colors refers to the graphical display of Junit, where passing tests are \emph{green} and failing tests are \emph{red}}.
This distinction does not consider the specification of exception handling.
Beyond green and red test cases, we characterize the test cases in three categories:
the pink, blue and white test cases.
Those three new types of test cases are a partition of passing test cases.

\paragraph{Pink Test Cases: Specification of Nominal Usage}
The ``\textcolor{magenta}{pink test cases}'' are those test cases where no exceptions at all are thrown or caught. The pink test cases specify the nominal usage of the software under test, i.e. the functioning of the system according to plan under standard input and environment. Note that a pink test case can still execute a try-block (but never a catch block by definition).

\paragraph{Blue Test Cases: Specification of State Incorrectness Detection}

Conceptually, there is an envelope that defines all possible correct states of an application. We call it the ``state correctness envelope".
This envelope is the boundary between correct and incorrect runtime states.
Delimiting the ``state correctness envelope'' can be achieved by writing test cases that simulate incorrect states, and then assert the presence in the test suite of exceptions of the expected type. 

The ``\textcolor{blue}{blue test cases}'' are those test cases which assert the presence of exception under incorrect input (such as for instance ``division(15,0)'').
The number of blue test cases $B$ estimates the amount of specification of the state correctness detection (by amount of specification, we mean the number of specified failure  scenarii).
$B$ is obtained by intercepting all bubbling exceptions, i.e. exceptions that quit the application code and arrive in the test case code.

\paragraph{White Test Cases: Specification of Exception Handling}

In our terminology, white test cases specify the required exception-handling capabilities of the system under test.
This specification is done  by 1) simulating the occurrence of an exception,
2) asserting that the exception is caught in application code and the system is in a correct state afterwards. 
If a test case still passes after the execution of a catch block in the application under test, it means that the recovery code in the catch block has successfully repaired the state of the program.

The ``\fcolorbox{white}{black}{\textcolor{white}{\textbf{white test cases}}}'' are those test cases that do not expect an exception (they are standard passing functional test cases) but use throw and catch at least once in application code.
Contrary to blue tests, they are not expecting thrown exceptions but they them only internally.

\begin{table}

\begin{tabularx}{\columnwidth}{l|X|p{2cm}|X}
 Test suite & \# tests & Blue tests \# (\%) & \# expected exceptions \\
\hline
 commons.lang & 2,046 & 393 (19\%)   & 817  \\
 commons.codec & 191 & 31 (16\%)   & 45  \\
 joda time & 3,950 & 656 (17\%) & 1,291  \\
 spojo core & 135 & 17 (13\%) & 17  \\
 sonar core & 100 & 6 (6\%) & 6  \\
 sonar plugin & 339 & 17 (5\%) & 17  \\
 jbehave core & 481 & 71 (15\%) & 99  \\
 shindig java gadgets & 2,031 & 154 (8\%) & 155  \\
 shindig common & 406 & 31 (8\%) & 132  \\
\end{tabularx}

\caption{The number of blue test cases. This shows that there is a specification of the detection capabilities of incorrect states in test suites.}
\label{tab:expected}
\vspace{-.4cm}
\end{table}

\subsection{Measuring the Amount of Specification of Error-Handling in Test Suites}

\textbf{Dataset: }
In this paper, we analyze the specification of error-handling in the test suites of 9 open-source projects:
Apache commons-lang, Apache commons-code, joda-time, Spojo core, Sonar core, Sonar Plugin, JBehave Core, Shindig Java Gadgets and Shinding Common.
The selection criteria are as follows.
First, the test suite has to be in the top 50 of most tested exceptions according to the SonarSource Nemo ranking\footnote{See \url{http://nemo.sonarsource.org}}.
SonarSource is the organization behind the software quality assessment platform ``Sonar''.
The Nemo platform is their show case, where open-source software is continuously tested and analyzed.
Second, the test suite has to be runnable within low overhead in terms of dependencies and execution requirements.

The line coverage of the test suites under study has a median of 81\%, a minimum of 50\% and a maximum of 94\%.

\paragraph{Proportion of Blue Test Cases}

\begin{table}

\begin{tabularx}{\columnwidth}{l|X|X|X}
 Test suite & \#test & White tests \# (\%)& thrown internal  \\
\hline
 commons.lang & 2,046 & 115  (6\%) & 1,106 \\
 commons.codec & 191 & 10  (5\%) &  16  \\
 joda time & 3,950 & 163  (4\%) &  337  \\
 spojo core & 135 & 8  (6\%) &  8  \\
 sonar core & 100 & 26  (26\%) &  168  \\
 sonar plugin & 339 & 12  (4\%) &  19  \\
 jbehave core & 481 & 94  (20\%) &  386  \\
 shindig java gadgets & 2031 & 172  (8\%) &  225  \\
 shindig common & 406 & 34  (8\%) &  143  \\
\end{tabularx}
\caption{The proportion of white test cases, i.e. test cases that do not expect exceptions but where exceptions are used internally by the application.}
\label{tab:internal-with-ex}
\vspace{-.4cm}
\end{table}

Table \ref{tab:expected} presents the proportion of blue test cases (those which expect exceptions under incorrect input).
The first and second columns are respectively the name of the application under analysis and the number of test cases in the corresponding test suite.
The  third column gives the absolute and relative proportion of blue test cases.
The fourth column is the number of expected exceptions (exceptions bubbling up to the test case).

One can see that between 5 and 19\% of test cases expect exceptions.
By construction, those blue test cases use at least one exception, but may use more than one.
Indeed, when comparing the third and the fourth columns (\# of blue tests versus number of expected exceptions), one sees that there are test cases that expect much more than one exception.
For instance Apache Shindig's test method \texttt{testDecryptGarbage} expects 100 exceptions.
Note that for many projects under study, the number of blue test cases is equal to the number of incoming exceptions. This indicates that the presence of a testing design rule: one expected exception per test.

\emph{Is the capability of detecting incorrect states specified?}
Yes, our results show that there exists a specification of the state correctness envelope.
The assertions of blue test cases specify both when an exception should be thrown and the type of the expected exception.
The number of specified exceptions is an approximation of the quantity of incorrect states anticipated and specified by the developers.

\paragraph{Proportion of White Test Cases}

To identify the white test cases (those which specify error-handling), we have set up the following experiment.
We run all test cases, trace the thrown exceptions and log those exceptions that do not bubble up to the test methods.
A test with at least one thrown exception and no bubbling ones is considered as white.

Table \ref{tab:internal-with-ex} gives the number and proportion of white test cases.
The second column recalls the number of test cases.
The third column gives the number of white test cases.
The fourth column indicates the number of exceptions involved in those white test cases.
All test suites expose white test cases.
In addition, three projects under study have more than 100 white test cases.

The proportion of pink test cases (with no exception at all) is the number of test cases that are neither blue nor white.
In our dataset, it varies between 65\% and 81\% of test cases.

\emph{Is the exception handling specified?}
Our results show that there exists specifications of exception handling as shown by the presence of a significant amount of white test cases.
This error-handling specification is necessary with several respects.
First, it gives data points for analyzing the actual behavior of error-handling with respect to specified inputs (this is what we do in section \ref{sec:verifying}).
Second, when one modifies the error-handling code with an automated approach, this specification enables one to check that the modified code still satisfy the specifed error-handling (this is what we do in section \ref{sec:improving}). 

\section{Verifying Software Resilience using Test Suites}
\label{sec:verifying}

\subsection{Definition of Two Contracts for Exception Handling}
\label{sec:contracts}

We now present two novel contracts for exception-handling programming constructs.
We use the term ``contract'' in its generic acceptation: 
a property of a piece of code that contributes to reuse, maintainability, correctness or another quality attribute.
For instance, the ``hashCode/equals'' contract\footnote{\url{http://docs.oracle.com/javase/7/docs/api/java/lang/Object.html\#hashCode()}} is a property on a pair of methods. 
This definition is broader in scope than Meyer's "contracts" \cite{meyer1992applying} which refer to preconditions, postconditions and invariants contracts.

We focus on contracts on the programming language construct try/catch, which we refer to as ``try-catch''.
A try-catch is composed of one try block and one catch block.
Note that a try with multiple catch blocks is considered as a set of pairs consisting of the try block and one of its catch blocks.
This means that a try with $n$ catch blocks is considered as $n$ try-catch blocks.
This concept is generalized in most mainstream languages, sometimes using different names (for instance, a catch block is called an ``except'' block in Python). In this paper, we ignore the concept of ``finally'' block \cite{JLS3:05:Sun} which is more language specific and much less used in practice \cite{cabral2007exception}.

\subsubsection{Source Independence Contract}

\begin{figure*}
\begin{minipage}[t]{.33\linewidth}
\begin{lstlisting}[frame=none,numbers=none,captionpos=t,caption={\sc An Example of Source-Independent Try-Catch Block.},label=fig:example-fault-inde-ok]
try{
 String arg = getArgument();
 String key = format(arg);
 return getProperty(key , isCacheActivated);
}catch(MissingPropertyException e){
  return "missing property";
}
 
\end{lstlisting}
\end{minipage}\textcolor{white}{xx}\begin{minipage}[t]{.33\linewidth}
\begin{lstlisting}[frame=none,numbers=none,captionpos=t,caption={\sc An Example of Source-Dependent Try-Catch Block.},label=fig:example-fault-inde-ko]
boolean isCacheActivated = false;
try{
  isCacheActivated = getCacheAvailability();
  return getProperty(key , isCacheActivated);
}catch(MissingPropertyException e){
  if( isCacheActivated ){
    return "missing property";
  }else{
    throw new CacheDisableException();
} }
\end{lstlisting}
\end{minipage}\textcolor{white}{xx}\begin{minipage}[t]{.33\linewidth}
\begin{lstlisting}[frame=none,numbers=none,captionpos=t,caption={\sc An Example of Purely-Resilient Try-Catch Block.},label=fig:example-pure-resi-ok]
try{
  return getPropertyFromCache(key);
}catch(MissingPropertyException e){
  return getPropertyFromFile(key);
} 
\end{lstlisting}
\end{minipage}
\vspace{-.4cm}
\end{figure*}

\textbf{Motivation: }
When a harmful exception occurs during testing or production, a developer has two possibilities.
One way is to avoid the exception to be thrown by fixing its root cause (e.g. by inserting a not null check to avoid a null pointer exception).
The other way is to write a try block surrounding the code that throws the exception.
The catch block ending the try block defines the recovery mechanism to be applied when this exception occurs.
The catch block responsibility is to recover from the particular encountered exception. 
By construction, the same recovery would be applied if another exception of the same type occurs within the scope of the try block at a different location.

This motivates the source-independence contract:
the normal recovery behavior of the catch block must work for the foreseen exceptions;
but beyond that, it should also work for exceptions that have not been encountered but may arise in a near future.

We define a novel exception contract, that we called ``source-independence''  as follows:
\begin{definition}
A try-catch is source-independent if the catch block proceeds equivalently, whatever the source of the caught exception is in the try block.
\end{definition}

For now, we loosely define ``proceeds equivalently'': if the system is still in error, it means that the error kind is the same; if the system has recovered, it means that the available functionalities are the same. Section \ref{sec:assesment} gives a formal definition.

For example, Listing \ref{fig:example-fault-inde-ok} shows a try-catch that \emph{satisfies} the source-independence contract.
If a value is missing in the application, and exception is thrown and the method returns a default value ``missing property".
The code of the catch block (only one return statement) clearly does not depend on the application state.
The exception can be thrown by any of the 3 statements in the try, and the result will still be the same.

On the contrary, Listing \ref{fig:example-fault-inde-ko} shows a try-catch that \emph{violates} the source-independence contract.
Indeed, the result of the catch process depends on the value of \emph{isCacheActivated}.
If the first statement fails (throws an exception), the variable \emph{isCacheActivated} is false, then an exception is thrown.
If the first statement passes but the second one fails, \emph{isCacheActivated} can be \emph{true}, then the value \emph{missing property} is returned.
The result of the execution of the catch depends on the state of the program when the catch begins (here it depends on the value of the \emph{isCacheActivated} variable).
In case of failure, a developer cannot know if she will have to work with a default return value or with an exception.
This catch is indeed source-dependent. 

\textbf{Discussion: }
How can it happen that developers write source-dependent catch blocks?
Developers discover exception risks due to a checked exception compilation verification, or at the first run-time occurrence of an unchecked exception.
In this case, the developer adds a try-catch block and puts the exception raising code in the try body.  
Often, the try body contains more code than the problematic statement in order to avoid variable scope and initialization problems.
However, while implementing the catch block, the developer still assumes that the exception can only be thrown by the ``problematic'' statement, and refers to variables that were set in previous statements. In other terms, the catch block is dependent on the application state at the problematic statement.
If the exception comes from the problematic statement, it works, if not, it fails.

We will present a formal definition of this contract and an algorithm to verify it in Section \ref{sec:assesment}.
We will show that both source-independent and source-dependent catch blocks exist in practice in Section \ref{sec:results-short-circuit}.

\subsubsection{Pure Resilience Contract }
\label{sec:resilience-contract}

\textbf{Motivation: }
In general, when an error occurs, it is more desirable  to recover from this error than to stop or crash.
A good recovery consists in returning the expected result despite the error and in continuing the program execution.

One way to obtain the expected result under error is to be able to do the same task in a way that, for the same input, does not lead to an error but to the expected result.
Such an alternative is sometimes called ``plan B''.
In terms of exception, recovering from an exception with a plan B means that the corresponding catch contains the code of this plan B.
The ``plan B" performed by the catch is an alternative to the ``plan A" which is implemented in the try block.
Hence, the contract of the try-catch block (and not only the catch or only the try) is to correctly perform a task T under consideration whether or not an exception occurs.
We refer to this contract  as the ``pure resilience" contract. 
 
A ``pure resilience'' contract applies to try-catch blocks.
We define it as follows: 

\begin{definition}
A try-catch is purely resilient if the system state is equivalent at the end of the try-catch execution whether or not an exception occurs in the try block.
\end{definition}

By system state equivalence, we mean that the effects of the plan A on the system are similar to those of plan B from a given observation perspective. If the observation perspective is a returned value, the value from plan A is semantically equivalent to the value of plan B (e.g. satisfies an ``equals'' predicate method in Java). We will precisely define this notion of equivalence using the specification given by test suites in Section \ref{sec:assesment}.

For example, Listing \ref{fig:example-pure-resi-ok} shows a purely resilient try-catch where a value is required, the program tries to access this value in the cache. If the program does not find this value, it retrieves it from a file.

\textbf{Usages: }
There are different use cases of purely resilient try-catch blocks.
We have presented in Listing \ref{fig:example-pure-resi-ok} the use case of caching for pure resilience.
One can use purely resilient try-catch blocks for performance reasons:
a catch block can be a functionally equivalent yet slower alternative. The efficient and more risky implementation of the try block is tried first, and in case of an exception, the catch block takes over to produce a correct result.
Optionality is another reason for pure resilience.
Whether or not an exception occurs during the execution of an optional feature in a try block,  the program state is valid and allows the execution to proceed normally after the execution of the try-block.

\textbf{Discussion: }
The difference between source-independence and pure resilience is as follows.
Source-independence means that \emph{under error} the try-catch has the same observable behavior.
In contrast, pure resilience means that \emph{in nominal mode and under error} the try-catch block has the same observable behavior.
This shows that pure-resilience subsumes source-independence: by construction, purely resilient catch blocks are source-independent.
The ``pure resilience" contract is a loose translation of the concept of recovery block \cite{horning1974program} in mainstream programming languages.

We will present a formal definition of this contract and an algorithm to verify it in Section \ref{sec:assesment}.

Although the ``pure resilience" contract is strong, we will show in Section \ref{sec:results-short-circuit} that we observe purely resilient try-catch blocks in reality, without any dedicated search: the dataset under consideration has been set up independently of this concern.

We presented the source independence and pure resilience contracts. Note that those contracts are not meant to be mandatory.
The try-catch blocks can satisfy one, both, or none.
We only argue that satisfying them is better from the viewpoint of resilience.

\subsection{An Algorithm and Four Predicates For Exception Contracts}
\label{sec:assesment}

We have defined in Section \ref{sec:contracts} two exceptions contracts applicable to try-catch blocks.
We now describe an algorithm and formal predicates to verify those contracts according to a test suite.
The algorithm collect data about the try-catch blocks (see Section \ref{sec:short-circuit}).
This data is then used to verify the predicates (see Section \ref{sec:predicates}).

\begin{algorithm*}
\KwIn{An Application $A$, A test suite $TS$ specifying the behavior of $A$.}
\KwOut{a matrix $M$ (try-catch $times$ test cases, the cells represent test success or failure.}
\Begin{
   $try\_catch\_list \gets static\_analysis( A )$\Comment{retrieve all the try-catch of the application}\\
   $standard\_behavior \gets standard\_run( TS )$\Comment{get test colors and try-catch behaviors}\\
       \For{$t \in try\_catch\_list$ \Comment{\em For each try-catch $t$ in the application}\\} {
          $prepare\_injection( t , c )$\Comment{prepare the try-catch $t$ by setting an injector}\\\Comment{ which will throw an exception of the type caught by $c$}\\\Comment{ at the beginning of each execution of the try $t$}\\
      $as \gets standard\_behavior:get\_test\_using( t )$  \Comment{retrieve all tests in the test suite $TS$}\\\Comment{ using the try of the try-catch $t$}\\
          \For{$a \in as$ \Comment{\em For all test $a$ which use the current try}\\} {
             $pass \gets run\_test\_with\_injection( a )$\Comment{get the result of the test under injection}\\
	      $M[ c , t ] = pass$\Comment{store the result of the catch $c$ for the test $t$ under injection}\\
          }
   }
   \textbf{return} $M$
}
\caption{The Short-Circuit Testing Algorithm. It uses exception injection to collect data about the behavior of catch blocks.}
\label{fig:algo}
\end{algorithm*}

\subsubsection{The Short-circuit Testing Algorithm}
\label{sec:short-circuit}

We now present a technique, called ``short-circuit testing'', which allows one to find source-independent and purely-resilient try-catch blocks. 
Short-circuit testing consists of injecting exceptions and then collecting execution data and test results.
This data is next analyzed to verify whether a try-catch block satisfies or violates the two contracts aforementioned.
This technique allows us to study the resilience of try-catch blocks in unanticipated scenarii.
We call this technique ``short-circuit testing'' because it ressembles electrical short-circuits: when an exception is injected, the code of the try block is somehow short-circuited. The name of software short-circuit has been introduced by the Hystrix resilience library\footnote{see \url{https://github.com/Netflix/Hystrix}}.

Injecting exceptions allows us to artificially create new system states.
Indeed, when one injects an exception in a try instead of letting it finish its execution, the system is put in an unexpected state.
So, we inject exceptions at appropriate places to simulate unanticipated errors.
In this way, we can observe new behaviors of this try-catch block.

What can we say  about a try-catch when a test passes while injecting an exception in it?
We use the test suite as an oracle of execution correctness:
if a test case passes under injection, the new behavior triggered by the injected exception is in accordance with the specification.
Otherwise, if the test case fails, the new behavior is detected as incorrect by the test suite.

Let us detail the behavior of this algorithm given in Figure~\ref{fig:algo}.
Our algorithm needs an application $A$ and its test suite $TS$. 

As defined in Section \ref{sec:contracts}, the contracts apply at the level of try-catch blocks.
First, a static analysis extracts the list of existing try-catch blocks.
For instance, the system extracts that method foo() contains one try with two associated catch blocks, they form two try-catch blocks.
In addition, we also need to know which test cases specifies which try-catch blocks.
Since we consider the test cases as specification, a test case is said to specify a  try-catch if it uses it.
To perform this, the first phase collects data about the standard run of the test suite under consideration.
For instance, the system learns that the try block in method foo() is executed on the execution of test \#1 and \#5.

This first run collects fine-grain data about the try-catch ``usages" defined as follows:
A ``pink usage" of a try block is when no exception is thrown;
A ``white usage" is when an exception is caught;
A ``blue usage" is when an exception is thrown but not caught.
Those color purposefully correspond to the test case colors defined in Section \ref{sec:specification-error-handling}.
A pink test does not execute any catch, in other terms its execution can only contains $pink usage$, or it does not execute any try-catch blocks.
A white test always executes at least one catch block. Hence, it contains at least one $white usage$.
A blue test necessarily throws an exception (by construction). If it is within the scope of a try block, there is at least one $blue usage$ or one $white usage$ (which may rethrow the exception). Otherwise, the thrown exception does not traverse any try block and there is no try-catch usages.
Note that a single try-catch can be executed multiple times, with different try-catch usages, in one single test.
This information is used later to verify the contracts (see Section~\ref{sec:predicates}).

Then, the algorithm injects exceptions during test suite execution and verifies that the test cases are still passing.
To assess the contracts at a try-catch level, the algorithm loops over the try-catch pairs (recall that a try with $n$ catch blocks is split into $n$ conceptual pairs of try/catch.
For each try-catch pair, the set of test cases using $t$, $a$, is already known thanks to monitoring.
It then executes each one of those tests while injecting an exception at the beginning of the corresponding try.

This simulates the worst-case exception, worst-case in the sense that it discards the whole code of the try block (it is future work to simulate exception at any possible locations).

Consequently, if the number of catch blocks corresponding to the executed try block  is N, there is one static analysis, one full run of the test suite and N runs of $as$.
In our example, the system runs its analysis, it executes the full test suite once.
Then it runs tests \#1 and \#5 with fault injection twice.
The first time the injection exception goes in the first catch block,
the second time, it goes, thanks to typing, in the second catch block.

The output of the algorithm is a matrix $M$ which represents the result of each test case under injection ( for each try-catch).
$M$ is a matrix of boolean values where each row represents a try-catch block, and each column represents a test case. A cell in the matrix indicates whether the test case passes with exception injection in the corresponding try-catch.

The exception contract predicates described next in Section~\ref{sec:predicates} are evaluated on $M$ that is obtained with short-circuit testing.
Short-circuit testing is performed with source code injection. Listing \ref{fault-injection} illustrates how this is implemented.
The injected code is able to throw an exception in a context dependent manner. The injector is driven by an exception injection controller at runtime.

\subsubsection{Resilience Predicates}
\label{sec:predicates}
We now describe four predicates that are evaluated on each row of the matrix to assess whether:
the try-catch is source-independent (contract satisfaction),
the try-catch is source-dependent  (contract violation),
the try-catch is purely-resilient (contract satisfaction),
the try-catch is not purely-resilient  (contract violation).

As hinted here, there is no one single predicate $p$ for which $contract[x] = p[x]$ and $\neg contract[x] = \neg p[x]$.
For both contracts, there are some cases where the short-circuit testing procedure yields not enough data to decide whether the contract is satisfied or violated. The law of the excluded third (\emph{principium tertii exclusi}) does not apply in our case.

\paragraph{Source Independence Predicate}

The decision problem is formulated as: given a try-catch and a test suite, does the source-independence contract hold?
The decision procedure relies on two predicates.

\textbf{Predicate \#1  ($source\_independent[x]$): Satisfaction of the source independence contract:}
A try-catch $x$ is source independent if and only if
 for all test cases that execute the corresponding catch block ($white\_usage$),
   it still passes when one throws an exception at the worst-case location in the corresponding try block. 

Formally, this reads as: 
\begin{align}
\label{for:src-indep}
\begin{split}
source\_independent [ x ] =  \forall a \in A_x |  \forall u_a \in usages\_in( x , a ) |  \\ ( is\_white\_usage[ u_a ] \implies pass\_with\_injection [ a , x ])
\end{split}
\end{align}

In this formula, $x$ refers to a try-catch (a try and its corresponding catch block), 
$A_x$ is the set of all tests executing $x$ (passing in the try block),
$u$ is a try-catch usage, i.e. a particular execution of a given try-catch block,
$usages\_in( x , a )$ returns the runtime usages of try-catch $x$ in the test case $a$,
$is\_white\_usage[u]$ evaluates to true if and only if an exception is thrown in the try block and the catch intercepts it,
$pass\_with\_injection$ evaluates to true  if and only if the test case $t$ passes with exception injection in try-catch $x$.

\textbf{Predicate \#2 ($source\_dependent[x]$):Violation of the source independence contract:}
A try-catch $x$ is not source independent if
 there exists a test case that executes the catch block which fails when one throws an exception at a particular location in the try block.

This is translated as:
\begin{align}
\label{for:not-src-indep}
\begin{split}
source\_dependent [ x ] =  \exists a \in A_x |  \forall u_a \in usages\_in( x , a ) |  \\ ( is\_white\_usage[ u_a ] \wedge \neg pass\_with\_injection [ a , x ])
\end{split}
\end{align}

\textbf{Pathological cases:} By construction $source\_dependent[x]$ and $source\_independent[x]$ cannot be evaluated to true at the same time (the decision procedure is sound).
If $source\_$\\$dependent[x]$ and $source\_independent[x]$ are both evaluated to false, it means that the procedure yields not enough data to decide whether the contract is satisfied or violated.

\begin{lstlisting}[float,captionpos=b,caption={Short-circuit testing is performed with source code injection. The injected code is able to throw an exception in a context dependent manner. The injector can be driven at runtime.},label=fault-injection]
try{
   // injected code
   if(Controller.isCurrentTryCatchWithInjection())
     if(Controller.currentInjectedExceptionType() == Type01Exception.class ){
        throw new Type01Exception();
     }else if(Controller.currentInjectedExceptionType() == Type02Exception.class ){
        throw new Type02Exception();
    }

   ... //normal try body
...
} catch (Type01Exception t1e) {
... //normal catch body
}  catch (Type02Exception t2e) {
... //normal catch body
} 
\end{lstlisting}

\paragraph{Pure Resilience Predicate}
The decision problem is formulated as: given a try-catch and a test suite, does the pure-resilience contract hold?
The decision procedure relies on two predicates.

\textbf{Predicate \#3  ($resilient[x]$): Satisfaction of the pure resilience contract}
A try-catch $x$ is purely resilient if 
it is covered by at least one pink test and
 all test cases that executes the try block pass when one throws an exception at the worst-case location in the corresponding try block.
In other words, this predicate holds when all tests pass even if one completely discards the execution of the try block.

Loosely speaking, a purely resilient catch block is a ``perfect plan B". 

This is translated as:
\begin{align}
\label{for:pure-resi}
\begin{split}
resilient [ x ] = (\forall a \in A_x  | pass\_with\_injection [ a , x ])\\
\wedge  ( \exists a \in A_x | \exists u_a \in usages\_in( x , a ) | is\_pink\_usage[ u_a ] )
\end{split}
\end{align}

where is\_pink\_usage[u] evaluates to true if and only if no exception is thrown in the try block.

\textbf{Predicate \#4  ($not\_resilient[x]$): Violation of the pure resilience contract}
A try-catch $x$ is not purely resilient if there exists a failing test case when one throws the exception at a particular location in the corresponding try block.

This predicate reads as:
\begin{align}
\label{for:not-pure-resi}
\begin{split}
not\_resilient[x] = \exists a \in A_c | \neg pass\_with\_injection [ a , x ] 
\end{split}
\end{align}

\textbf{Pathological cases} By construction $resilient[x]$ and $not\_$\\$resilient[x]$ cannot be evaluated to true at the same time (the decision procedure is sound).
Once again, if they are both evaluated to false, it means that the procedure yields not enough data to decide whether the contract is satisfied or violated.

\section{Improving Software Resilience with Catch Stretching}
\label{sec:improving}

\begin{table*}
\begin{tabularx}{\textwidth}{l|X||X|X|X||X|X||X}
 & \# executed try-catch  & \# purely resilient try-catch &\# source-independent try-catch &\# source-dependent try-catch & Unknown w.r.t. resilience & Unknown w.r.t. source independence & \# Stretchable try-catch\\
\hline
 commons-lang & 49 & 3/49 & 18/49 & 5/49 & 1/49 & 26/49 & 16/18 \\
 commons-codec & 14 & 0/14 & 12/14 & 0/14 & 0/14 & 2/14 & 12/12\\
 joda time & 18 & 0/18 & 4/18 & 0/18 & 2/18 & 14/18 & 4/4\\
 spojo core & 1 & 1/1 & 1/1 & 0/1 & 0/1 & 0/1 & 1/1\\
 sonar core & 10 & 0/10 & 9/10 & 1/10 & 1/10 & 0/10 & 7/9\\
 sonar plugin  & 6 & 0/6 & 3/6 & 0/6 & 0/6 & 3/6 & 3/3\\
 jbehave core & 42 & 2/42 & 7/42 & 2/42 & 9/42 & 33/42 & 7/7\\
 shindig-java-gadgets & 80 & 2/80 & 30/80 & 12/80 & 21/80 & 38/80 & 26/30\\
 shindig-common & 21 & 1/21 & 8/21 & 4/21 & 4/21 & 9/21 & 8/8\\
\hline
 total & 241 & 9 & 92 & 24 & 38 & 125 & 84/92\\
\end{tabularx}
\caption{The number of source independent , purely resilient and stretchable catch blocks found with short-circuit testing. Our approach provides developers with new insights on the resilience of their software.}
\label{THE-table}
\vspace{-.4cm}
\end{table*}

We know that some error-handling is specified in test suites (Section \ref{sec:specification-error-handling}).
We have defined two formal criteria of software resilience and an algorithm to verify them (Section~\ref{sec:assesment}).
How to put this knowledge in action?

For both contracts, one can improve the test suite itself.
As discussed above, some catch blocks are never executed and others are not sufficiently executed to be able to infer their resilience properties (the pathological cases of Section~\ref{sec:assesment}).
The automated refactoring of the test suite is outside the scope of this paper
but we will discuss in Section~\ref{sec:results-short-circuit} how developers can manually refactor their test suites to improve the automated reasoning on the resilience.
\\
\\

\paragraph{Definition of Catch Stretching}
We now aim at improving the resilience against unanticipated exceptions, those exceptions that are not specified in the test suite and even not foreseen by the developers. According to our definition of resilience set in the introduction, this means improving the capability of the software under analysis to correctly handle unanticipated exceptions.
For this, a solution is to transform the catch so that they catch more exceptions than before.
This is what we call ``catch stretching'': replacing the type of the caught exceptions.
For instance, replacing \texttt{catch(FileNotFoundException e)} by \texttt{catch(IO\-Exception e)}.
The extreme of catch stretching is to para\-metrize the catch with the most generic type of exceptions (e.g. \texttt{Throwable} in Java, \texttt{Exception} in .NET).
This transformation may look naive, but there are strong arguments behind it.
Let us now examine them.

We claim that \emph{all source-independent catch blocks are candidates to be stretched}. This encompasses purely-resilient try-catch blocks as well since by construction they are also source-independent (see Section \ref{sec:contracts}).
The reasoning is as follows.

\paragraph{Catch Stretching Under Short-Circuit Testing}

By stretching source independent catch-blocks, the result is equivalent under short-circuit testing.
In the original case, injected exceptions are of type $X$ and caught by  \texttt{catch(X e)}.
In the stretched case, injected exceptions are of generic type $Exception$ and caught by  \texttt{catch(Exception e)}.
In both cases, the input state of the try block is the same (as set by the test case), 
and the input state of the catch block is the same (since no code has been executed in the try block due to fault injection).
Consequently, the output state of the try-catch is exactly the same.
Under short-circuit testing, catch stretching yields strictly equivalent results.

\paragraph{Catch Stretching and Test Suite Specification}
\label{sec:tests-stretch}

Let us now consider a standard run of the test suite and a source-independent try-catch.
In standard mode, with the original code, there are two cases: either all the exceptions thrown in the try block under consideration are caught by the catch (case A), or at least one traverses the try block without being caught because it is of an uncaught  type (case B). In both cases, we refer to exceptions normally triggered by the test suite, not injected ones.

In the first case, catch stretching does not change the behavior of the application under test: all exceptions that were caught in this catch block in the original version are still caught in the stretched catch block. In other words, the stretched catch is still correct according to the specification. And it is able to catch many more unanticipated exceptions: it corresponds to our definition of resilience. On those source-independent try-catch of case (A), \emph{catch stretching improves the resilience of the application}.

We now study the second case (case B): there is at least one test case in which the try-catch $x$ under analysis is traversed by an uncaught exception.
There are again two possibilities: this uncaught exception bubbles to the test case, which is a blue test case (the test case specifies that an exception must be thrown and asserts that it is actually thrown).
If this happens, we don't apply catch stretching.
Indeed, it is specified that the exception must bubble, and to respect the specifications we must not modify the try-catch behavior.
The other possibility is that the uncaught exception in try-catch $x$ is caught by another try-catch block $y$ later in the stack.
By definition, this corresponds to white test cases.
When stretching try-catch $x$, one replaces the recovery code executed by try-catch $y$ by executing the recovery code of try-catch $x$. 
However, it may happen that the recovery code of $x$ is different from the recovery code of $y$, and that consequently, the test case that was passing with the execution of the catch of $y$ (the original mode) fails with the execution of the catch $x$.

To overcome this issue, we propose to again use the test suite as the correctness oracle.
For source-independent try-catch blocks of case B, one stretches the catch to ``Exception'', one then runs the test suite, and if all tests still pass, we keep the stretched version. As for case A, the stretching enables to handle more unanticipated exceptions while remaining correct with respect to the specification. Stretching source-independent try-catch blocks of both case A and case B improves the resilience.

\paragraph{Summary}
To sum up, improving software resilience with catch stretching consists of:
First, stretching all source-independent try-catch blocks of case A.
Second, for each source-independent try-catch blocks of case B, running the test suite after stretching to check that the transformation has produced correct code according to the specification.
Third, running the test suite with all stretched catch blocks to check whether there is no strange interplay between all exceptions.

We will show in Section \ref{sec:results-stretching}  that most (91\%) of source-independent try-catch blocks can be safely stretched according to the specification.

\section{Empirical Evaluation}
\label{sec:results-short-circuit}

We have presented two exception contracts: pure resilience and source independence (Section \ref{sec:resilience-contract}).
We now evaluate those contracts from an empirical point of view.
Can we find real world try-catch blocks for which the corresponding test suite enables us to prove their source independence? Their pure resilience capability?

The experimental protocol is as follows.
We run the short-circuit testing algorithm described in Section~\ref{sec:short-circuit} on the 9 reference test suites described in Section~\ref{sec:specification-error-handling}.
As seen in Section~\ref{sec:short-circuit}, short-circuit testing runs $n$ times the test suite, where $n$ is the number of executed catch blocks in the application.
In total, we have thus 241 executions over the 9 test suites of our dataset.

Table~\ref{THE-table} presents the results of this experiment.
For each project of the dataset and its associated test suite, it gives 
the number of executed catch blocks during the test suite execution, 
purely resilient try-catch blocks,
source-independent try-catch blocks,
and the number of try-catch blocks for which runtime information is not sufficient to assess the truthfulness of our two exception contracts.

\paragraph{Source Independence}
Our approach is able to demonstrate that 92 try-catch blocks (sum of the fourth column of Table~\ref{THE-table}) are source-independent (to the extent of the testing data).
This is worth noticing that with no explicit ways for specifying them and no tool support for verifying them, some developers still write catch blocks satisfying this contract.
This shows that our contracts are not theoretical: they cover a reality of error-handling code.

Beyond this, the developers not only write some source-independent catch blocks, they also write test suites that provide enough information to decide with short-circuit testing whether the catch is source-independent or not.

Our approach also identifies 24 try-catch blocks that are source-dependent, i.e. that violate the source-independence predicate.
Our approach makes the developers aware that some catch blocks are not independent of the source of the exception: \emph{the catch block implicitly assumes that the beginning of the try has always been executed when the exception occurs}.
Within the development process, this is a warning.
The developers can then fix the try or the catch block if they think that this catch block should be source independent or choose to keep them source-dependent, in total awareness.
It is out of the scope of this paper to automatically refactor source-dependent try-catch blocks as source-independent.

For instance, a source-dependent catch block  of the test suite of sonar-core is shown in Listing \ref{sonar-dependent}.
Here the "key" statement is the \emph{if (started == false)} (line 6).
Indeed, if the call to \emph{super.start()} throws an exception before the variable \emph{started} is set to true (\emph{started = true} line 15), an exception is thrown (line 7).
On the contrary, if the same DatabaseException occurs after line 15, the catch block applies some recovery by setting default value (\emph{setEntityManagerFactory}).
Often, source-dependent catch blocks contain if/then constructs.
To sum-up, short-circuit testing catches assumptions made by the developers, and uncover causality effects between the code executed within the try block and the code of the catch block.

Finally, our approach highlights that, for 24 catch blocks (fifth column of Table~\ref{THE-table}), there is not enough tests to decide whether the source-inde\-pendence contract holds.
This also increases the developer awareness. 
This signals to the developers that the test suite is not good enough with respect to assessing this contract. This knowledge is directly actionable: for assessing the contracts, the developer has to write new tests or refactor existing ones.
In particular, as discussed above, if the same test case executes several times the same catch block, this may introduce noise to validate the contract or to prove its violation.
In this, the refactoring consists of splitting the test case so that the try/catch block under investigation is executed only once.

\begin{lstlisting}[float,captionpos=b,caption={ A Source-Dependent Try-Catch Found in Sonar-core using Short Circuit Testing.},label=sonar-dependent]
public class MemoryDatabaseColector extends AbstractDatabaseColector {
   public void start(){
      try{
         super.start(); // code below
      }catch (DatabaseException ex) {
          if (started==false) // this is the source-dependence
              throw ex;
          setEntityManagerFactory();
      }}}

 public void start(){
    ...  
    // depending on the execution of the following statement
    // the catch block of the caller has a different behavior 
    started = true;
    ...}}
\end{lstlisting}

\paragraph{Pure Resilience}
We now examine the pure-resilience contracts.
In our experiment, we have found 9 purely resilient try-catch blocks in our dataset.
The distribution by application is shown in the third column of Table~\ref{THE-table}.

Listing~\ref{fig:pure-resi-spojo} shows a purely resilient try-catch block found in project spojo-core using short-circuit testing.
The code has been slightly modified for sake of readability.
The task of the try-catch block is to return an instantiable Collection class which is compatible with the class of a prototype object.
The plan A consists of checking that the class of the prototype object has an accessible constructor (simply by calling \texttt{getDeclaredConstructor}). If there is no such constructor, the method call throws an exception.
In this case, the catch block comes into rescue and chooses from a list of known instantiable collection classes one that is compatible with the type of the prototype object.
According to the test suite, the try-catch is purely resilient: always executing plan B yields passing test cases.

The pure resilience is much stronger than the source independence contract.
While the former states that the catch has the same behavior wherever the exception comes from, the latter states that the correctness as specified by the test suite is not impacted in presence of unanticipated exceptions.
Consequently, it is normal to observe much less try-catch blocks verifying the pure resilience contract compared to the source-independent contract.
Despite the strength of the contract, this contract also covers a reality:
perfect alternatives, ideal plans B exist in real code.

One also sees that there are some try-catch blocks for which there is not enough execution data to assess whether they are purely resilient or not. This happens when a try-catch is only executed in white try-catch usages and in no pink try-catch usage. By short-circuiting the white try-catch usages (those with internally caught exceptions), one proves it source-independence, but we also need to short-circuit a nominal pink usage of this try-catch to assess that  plan B (of the catch block) works instead of  plan A (of the try block). This fact is surprising: this shows that some try-catch blocks are only specified in error mode (where exceptions are thrown) and not in nominal mode (with the try completing with no thrown exception). This also increases the awareness of the developers: for those catch blocks, test cases should be written to specify the nominal usage.

\begin{lstlisting}[float,captionpos=b,caption={ A Purely-Resilient Try-Catch Found in spojo-core (see SpojoUtils.java)},label=fig:pure-resi-spojo]
// task of try-catch: 
// given a prototype object
Class clazz = prototype.getClass();
// return a Collection class that has an accessible constructor
// which is compatible with the prototype's class
try {
	// plan A: returns the prototype's class if a constructor exists		
	prototype.getDeclaredConstructor();
	return clazz;
} catch (NoSuchMethodException e) {
	// plan B: returns a known instantiable collection
	// which is compatible with the prototype's class
	if (LinkedList.class.isAssignableFrom(clazz)) {
		return LinkedList.class;
	} else if (List.class.isAssignableFrom(clazz)) {
		return ArrayList.class;
	} else if (SortedSet.class.isAssignableFrom(clazz)) {
		return TreeSet.class;
	} else {
		return LinkedHashSet.class;
	}
}
\end{lstlisting} 

\paragraph{Catch Stretching}
\label{sec:results-stretching}

We look at whether, among the 92 source-independent try-catch blocks of our dataset, we can find stretchable ones (stretchable in the sense of Section \ref{sec:improving}, i.e. for which the caught exception can be set to ``Exception'').
We use source code transformation and the algorithm described in Section~\ref{sec:improving}.

The last column of Table~\ref{THE-table} gives the number of stretchable try-catch blocks out of the number of source-independent try-catch blocks. For instance, in commons-lang, we have found 18 candidates source-independent try-catch blocks. Sixteen (16/18) of them can be safely stretched: all test cases pass after stretching.

Table~\ref{THE-table} indicates two results. First, most (91\%) of the source-independent try-catch blocks can be stretched to catch all exceptions. In this case, the resulting transformed code is able to catch more unanticipated exceptions while remaining correct with respect to the specification. 

Second, there are also try-catch blocks for which catch stretching does not work.
As explained in Section \ref{sec:improving}, this corresponds to the the case where the stretching results in hiding correct recovery code (w.r.t. to the specification), with new one (the code of the stretched catch) that proves unable to recover from a traversing exception.

In our dataset, we encounter all cases discussed in Section \ref{sec:improving}. 
For instance in joda-time, all four source-independent try-catch blocks represent are never traversed by an exception -- case A of Section \ref{sec:tests-stretch}.
(for instance the one at line 560 of class ZoneInfoCompiler). 
We have shown that analytically, they can safely stretched. We have run the test suite after stretching, all tests pass. 

We have observed the two variations of case B (try-catch blocks traversed by exceptions in the original code). 
For instance, in sonar-core, by stretching a NonUniqueResultException catch to the most generic exception type, an IllegalStateException is caught.
However, this is an incorrect transformation, it results in one failing test case.

Finally, we discuss the last and most interesting case.
In commons-lang, the try-catch at line 826 of class ClassUtils can only catch a ClassNotFoundException but is traversed by a NullPointerException during the execution of the test ClassUtilsTest.testGetClassInvalidArguments.
By stretching ClassNotFoundException to the most generic exception type, the  NullPointerException is caught: the catch block execution replaces another catch block upper in the stack. 
Although the stretching modifies the test case execution, the test suite passes, this means that the stretching is correct with respect to the test suite.

\paragraph{Summary}

To sum up, this empirical evaluation has shown that the short-testing approach of exception contracts enables to increase the knowledge one has on a piece of software.
First, it indicates source-independent and purely resilient try-catch blocks. This knowledge is actionable: those catch blocks can be safely stretched to catch any type of exceptions. 
Second, it indicates source-dependent try-catch blocks. This knowledge is actionable: it says that the error-handling should be refactored so as to resist to unanticipated errors. 
Third, it indicates ``unknown'' try-catch blocks. This knowledge is actionable: it says that the test suite should be extended and/or refactored to support automated analysis of exception-handling.

\section{Related work}
\label{sec:rw}

Segal et al. \cite{segall1988fiat,barton1990fault} invented Fiat,  an early validation system based on fault injection.
Their fault model simulates hardware fault (bit changes in memory).
Kao et al. \cite{Kao1993} have described ``Fine'', a fault injection system for Unix kernels. It simulates both hardware and operating system software faults.
In comparison, we inject high-level software faults (exceptions) in a modern platform (Java).  
Bieman et al. \cite{bieman1996using} added assertions in software that can be handled with an ``assertion violation'' injector.
The  test driver enumerates different state changes that violate the assertion.
By doing so, they are able to improve branch coverage, especially on error recovery code.
This is different from our work since: we do not manually add any information in the system under study  (tests or application).
Fu et al. \cite{fu2003compiler}  described a fault injector for exceptions similar to ours in order to improve catch coverage.
In comparison to both \cite{bieman1996using} and \cite{fu2003compiler},
we do not aim at improving the coverage but to identify the try-catch blocks satisfying exception contracts.

Sinha \cite{sinha2000analysis} analyzed the effect of exception handling constructs (throw and catch) on different static analyses.
In contrast, we use dynamic information for reasoning on the exception handling code. 
The same authors described \cite{sinha2004automated} a complete tool chain to help programmers working with exceptions.
The information we provide (the list of source-independent, purely-resilient try-catch blocks and so forth) is different, complementary and may be subject to be integrated in such a tool.
 
Candea et al. \cite{candea2003automatic} used exception injection to capture the error-related dependencies between artifacts of an application. They inject checked exceptions as well as 6 runtime, unchecked exceptions.
We also use exception injection but for a different goal: verifying try-catch contracts.

Ohe et al. \cite{ohe2005exception} described an exception monitoring system that resembles ours. Beyond the monitoring system we also provide a strategy and a set of analyses to verify two exception contracts.

Ghosh and Kelly \cite{ghosh2008bytecode} did a special kind of mutation testing for improving test suites.
Their fault model comprises ``abend'' faults: abnormal ending of catch blocks. It is similar to short-circuiting.
We use the term ``short-circuit'' since it is a precise metaphor of what happens. In comparison, the term ``abend'' encompasses many more kinds of faults.   
In our paper, we claim that the new observed behavior resulting from short-circuit testing should not be considered as mutants to be killed.
Actually we claim the opposite: short-circuiting should remain undetected for sake of source independence and pure resilience. 

Fu and Ryder \cite{Fu2007} presented a static analysis for revealing the exception chains (exception encapsulated in one another). In contrast, our approach is a dynamic analysis.
We do not focus on exception chains, we propose an analysis of source-independence and pure resilience.
Mercadal \cite{mercadal2010} presented an approach to manage error-handling in a specific domain (pervasive computing).
This is forward engineering. On the contrary, we reason on arbitrary legacy Java code, we identify resilient locations and modifies others.

Zhang and Elbaum \cite{zhang2012amplifying} have recently presented an approach that amplifies test to validate exception handling.
Their work has been a key source of inspiration for ours.
Short-circuit testing is a kind of test amplification.
While the technique is the same, the problem domain we explore is really different.
They focus on exceptions related to external resources.
We focus on any kind of exceptions in order to verify resilience contracts.

\section{Conclusion}

In this paper, we have explored the concept of software resilience against exceptions.
We have contributed with different results that, to our knowledge, are not discussed in the literature.
First, we have shown to what extent test suites specify exception-handling.
Second, we have formalized two formal resilience properties: source-independence and pure-resilience as well as an algorithm to verify them.
Finally, we have proposed a source code transformation called ``catch stretching'' that improves the ability of the application under analysis to handle unanticipated exceptions.
Our future work consists in, first extending short-circuit to inject exceptions at any location of try-blocks and second, in exploiting this information to inject catch blocks at new places.
In particular, the purely resilient catch blocks have a real recovery power that could probably be used elsewhere.
  
\newpage
\bibliographystyle{abbrv}
\bibliography{va2cd242bab334a26}
\end{document}